\documentclass[preprint]{elsarticle}

\journal{Physics Letters B}

\usepackage{color,soul}
\usepackage{graphicx}
\usepackage{amsmath}

\newcommand {\la} {\langle}\newcommand {\ra} {\rangle}
\newcommand {\beq} {\begin{eqnarray}}
\newcommand {\eeq} {\end{eqnarray}}
\newcommand {\eeqn} [1] {\label{#1} \end{eqnarray}}
\newcommand {\eol} {\nonumber \\}
\newcommand {\ve} [1] {\mbox{\boldmath $#1$}}

\begin{document}

\begin{frontmatter}

\title{Single-particle spectroscopic strength from nucleon transfer reactions with a three-nucleon force contribution}

\author[surrey]{N.K. Timofeyuk\corref{cor1}}
\ead{n.timofeyuk@surrey.ac.uk}
\author[surrey]{L. Moschini}
\ead{l.moschini@surrey.ac.uk}
\author[seville]{M. G\'omez-Ramos}
\ead{mgomez40@us.es}

\cortext[cor1]{Corresponding author}
\address[surrey]{Department of Physics, Faculty of Engineering and Physical Sciences, University of Surrey, Guildford, Surrey GU2 7XH, United Kingdom}
\address[seville]{Departamento de FAMN, Universidad de Sevilla, Apartado 1065, 41080 Sevilla, Spain}


\date{\today}

\begin{abstract}
The direct reaction theory widely used to study single-particle spectroscopic strength in   nucleon transfer experiments is based on a Hamiltonian with two-nucleon   interactions only. We point out that in reactions with a loosely-bound projectile, where clustering and breakup effects are important, an additional  three-body force arises due to three-nucleon ($3N$) interaction between two nucleons belonging to  different clusters in the projectile and  a target nucleon. We study the effects of this force on nucleon transfer  in $(d,p)$ and $(d,n)$ reactions on $^{56}$Ni, $^{48}$Ca, $^{26m}$Al and $^{24}$O targets at deuteron incident energies between 4 and 40 MeV/nucleon. Deuteron breakup is treated exactly within a continuum discretized coupled-channel approach. It was found that an additional three-body force can noticeably alter the angular distributions at forward angles, with consequences for spectroscopic factors' studies.
Additional study of transfer to $2p$ continuum in the $^{25}$F$(p,2p)^{24}$O reaction,   involving the same overlap function as in the $^{24}$O($d,n)^{25}$F case,  revealed that $3N$ force  affects   the $(d,n)$ and $(p,2p)$ reactions in a similar way, increasing the cross sections  and  decreasing spectroscopic factors, although its influence at the main peak of $(p,2p)$  is weaker. The angle-integrated cross sections are found to be less sensitive to the 3N force contribution, they increase by less than  20\%.   Including 3N interactions in nucleon removal reactions makes an essential step towards bringing together nuclear structure theory, where 3N force is routinely used, and nuclear direct reaction theory, based on two-nucleon interactions only.
\end{abstract}

\end{frontmatter}

The spectroscopic strength of nucleon states in atomic nuclei, often associated with spectroscopic factors, is central to our understanding of nuclear structure.  Due to its connections with nucleon orbit occupancy \cite{Mac60} it has received enormous attention for the last 60 years from  the nuclear structure community. Today, the rapid progress of ab-initio treatments of nuclear structure \cite{Her20} has enabled spectroscopic factors to be related to realistic forces between nucleons.

Spectroscopic factors are often determined from nucleon transfer and nucleon removal experiments by comparing measured and calculated cross sections. Over the last two decades experimental studies scrutinized spectroscopic factor uncertainties using different reaction probes,  at the same time extending significantly the range of studied isotopes thanks to the increased availability of radioactive beams worldwide.
The experimental studies revealed that spectroscopic factors can be significantly lower than structure model predictions  even for double magic closed shell nuclei \cite{Kra01}. This phenomenon,   named  ``spectroscopic-factor quenching", occurs because  nucleon-nucleon (NN) correlations  scatter nucleons beyond the mean-field shell model space. The most puzzling discovery from a remarkable set of data collected over two decades is the  quenching dependence on neutron-proton binding asymmetry seen in  inverse-kinematic nucleon knockout experiments with $^9$Be target  \cite{Tos21}  and the absence of significant asymmetry-dependent quenching in nucleon transfers, such as $(p,d)$ reactions \cite{Fla18,Man21,Aum21}. Given that spectroscopic-factor determination  heavily relies on reaction theory, the nuclear physics community agrees that nucleon-removal reaction models should be further developed and, in particular, moved towards a better integration and coherent description with modern nuclear structure theories. However, the challenges in this direction are significant. 

In this Letter, we present a new step towards integrating nuclear reaction and structure theories by pointing out that analysis of all  nucleon transfer and nucleon removal  experiments is carried out using distorted-wave-type direct reaction models   based on a Hamiltonian with NN interaction only \cite{Satchler}. However, it has been known since the 1950s  that the three-nucleon (3N) force is important for the correct description of atomic nuclei. In reactions with loosely-bound projectiles, where clustering and breakup effects are important,  the $3N$ interaction between two nucleons belonging to  different clusters in the projectile and a target nucleon creates an additional three-body force \cite{Tim20}. We aim  to understand if this novel force can give any noticeable effect on nucleon transfer cross sections so that
that could eventually contribute to resolving the puzzling contradiction in spectroscopic factors quenching  deduced from  transfer and knockout experiments.
We choose the deuteron as an example of a loosely-bound projectile and consider $(d,p)$ and $(d,n)$ 
reactions - a popular tool choice for spectroscopic factor studies. 
Due to the  small binding energy of 2.2 MeV the 
deuteron breaks up easily into neutron $n$ and proton $p$
when interacting with a target, invoking a need for a three-body treatment of 
its motion before transfer.  
In addition, 
we  study one case of a popular alternative to   knockout experiments with heavy ions - $(p,2p)$ -  where the two final protons are in the continuum and the three-body description of the final state is mandatory.


We consider an $n$-$p$-$A$ model for $d$-$A$ scattering, where $A$ is the target, taking the $n$-$A$ and $p$-$A$ interactions $U_{nA}$ and $U_{pA}$ from  optical model,
and add  to the three-body Hamiltonian a 3N-induced contribution $\la \phi_A | \sum_j W_{jnp}| \phi_A\ra$, where $\phi_A$ is the target wave function and $W_{jnp}$ is the 3N interaction of target nucleon $j$ with neutron $n$ and proton $p$ from deuteron. We  solve the $n$-$p$-$A$ problem in the continuum-discretized coupled-channel (CDCC) approach \cite{Raw74,Aus87}. 
This requires constructing continuum bins $\phi_i$ from $n$-$p$ scattering wave functions and calculating the matrix elements
$U^{(ii')}_{3N}(\ve{R}) = \la \phi_i\phi_A | \sum_j W_{jnp}| \phi_A \phi_{i'}\ra$ (or coupling potentials) which are function of the coordinate $\ve{R}$ connecting the $n$-$p$ centre of mass with the target $A$. The radial part of these matrix elements could be read into a CDCC reaction code, which in our case was FRESCO \cite{FRESCO}, to supplement the coupling potentials $U^{(ii')}_{\rm opt}(\ve{R}) = $ $\la\phi_A \phi_i| U_{pA}+U_{nA} | \phi_A \phi_{i'}\ra$ arising from optical $n$-$A$ and $p$-$A$ interactions.
To develop $U^{ii'}_{3N}(\ve{R})$, we first calculate the effective  transition interaction $\tilde W^{(ii')}_{dj}=\la\phi_i^{I_d m_d}| W_{jnp} |\phi_{i'}^{I'_d m'_d}\ra$  connecting initial and final bin states of the $n$-$p$ pair with the nucleon $j$ belonging to the target $A$. We
 express it using a standard partial wave decomposition:
\beq
\la\phi_i^{I_d m_d}| W_{jnp} |\phi_{i'}^{I'_d m'_d}\ra &=& \sqrt{4\pi}
\sum_{\lambda I_d I'_d} {\tilde W}^{(ii')}_{\lambda I_d I'_d}(r_{dj})
(\lambda \mu I'_d m'_d | I_d m_d) Y^*_{\lambda \mu}(\hat{\ve{\rm r}}_{di}),
\eeqn{partial_decomposition}
where $I_d$ and $m_d$ are the total angular momentum and its projection of bin $i$, the coordinate ${r}_{di}$ connects the $n$-$p$ centre of mass and target nucleon $j$ and  $Y$ is the spherical function. 
The $d$-$j$ interaction is then folded with the target density $\rho_A$  to obtain the radial part $U^{(ii')}_{\lambda I_d I'_d}(R)$ of the coupling potential  $U^{ii'}_{3N}(\ve{R})$:
\beq
U^{(ii')}_{\lambda I_d I'_d}(R)
&=& (-)^{I_d-I'_d}4\pi (2\lambda+1)(2 {I}'_d +1)
\int_0^{\infty} dr_{dj} r_{dj}^2\rho_{\lambda}(r_{dj},R){\tilde W}^{(ii')}_{\lambda I_dI'_d}(r_{dj}), \eol
\eeqn{Ufresco}
where it was assumed that the target $A$ is spherical so that
\beq
\rho_{\lambda}(r_{dj},R) 
  = \frac{1}{2} \int_{-1}^1 d\mu \, P_{\lambda}(\mu) \rho_A \left( \left|\ve{r}_{dj} - \ve{R} \right| \right), \,\,\,\,\,\,\,\,
\eeqn{}
and $P_{\lambda}(\mu)$ are the Legendre polynomials.

The  effective potential $\tilde W^{(ii')}_{dj}$ 
is determined by the strength of  $W_{jnp}$ and by the
$n$-$p$ wave functions behaviour within its short range, thus requiring a consistent choice of the NN and 3N interactions. 
Today, the most popular consistent choice of NN+3N interactions is provided by the chiral effective field theory ($\chi$EFT) \cite{Mac11}. However, we are not yet in a position to use it because most $\chi$EFT potentials are nonlocal while the CDCC can only handle local interactions. In addition, the spin-isospin structure of the $\chi$EFT operators (both nonlocal and a few available local ones) requires knowledge of the spin density distributions in the target $A$, which is not yet available.
Therefore, we choose a phenomenological NN force AV18 \cite{AV18} supplemented by Urbana IX (UIX) interaction  \cite{Pud95} - a combination  successfully used in ab-initio Green's Function Monte Carlo calculations of light nuclei \cite{Pud97} and scattering studies in few-body systems \cite{Kie95,Viv96,Viv98}. The UIX potential consists of a contact and $2\pi$-exchange terms and   
the formal derivation of (rather cumbersome) expressions for $\tilde W^{(ii')}_{dj}$ involving these contributions is given in Supplement for   targets with spin $0^+$.

The following zero-spin targets are considered here: $^{48}$Ca, $^{56}$Ni, $^{26m}$Al and $^{24}$O. The densities of the first three of them were taken from a compilation \cite{deV87} of charge densities extracted from electron scattering. We assumed   that    symmetrical $N=Z$ nuclei $^{56}$Ni and $^{26m}$Al have the same  proton and nucleon density distributions  while in $^{48}$Ca they are proportional to each other. For $^{24}$O three different models were used: 
renormalised $^{18}$O harmonic-oscillator  from \cite{deV87}, the $^{24}$O Hartree-Fock    with SkM interaction and the SkM-Hartree-Fock $^{22}$O-core density plus valence-nucleon-density from Gamow shell model  of \cite{Li21}.
For all four isotopes different choices of the target density did not affect the final cross sections. 

\begin{figure}[t]
\includegraphics[scale=0.33]{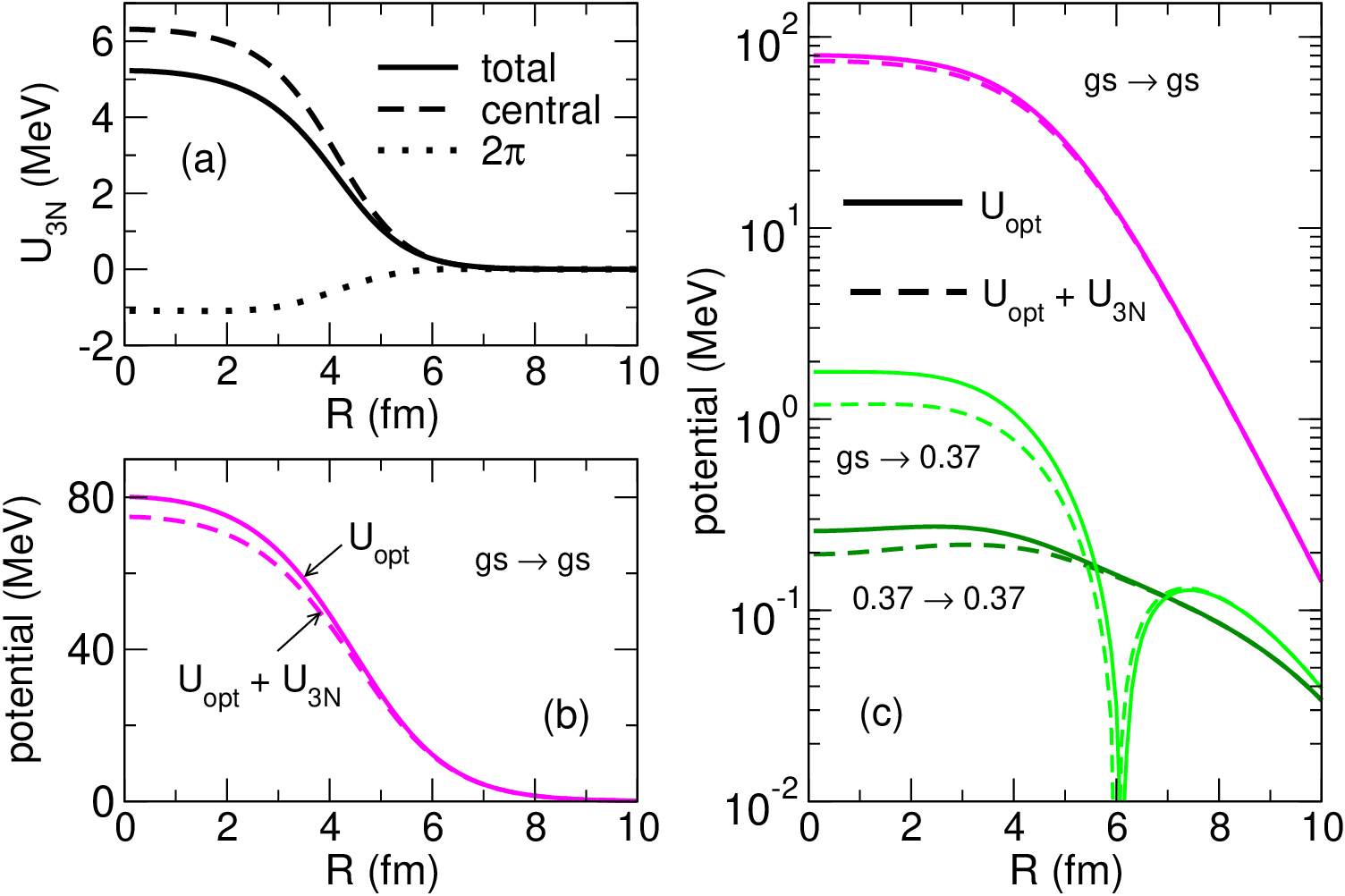}
\caption{
$(a)$  $U_{3N}^{00}$ potential for $d+^{56}$Ni calculated with contact (dashed line), $2\pi$-exchange (dotted line) contributions and their sum (solid line). (b) Absolute values of $U_{\rm opt}^{ii'}$   (solid lines) and its sum with $U_{3N}^{ii'}$ (dashed lines) for $i=i'=0$, (c) the same as (b) for two diagonal and one non-diagonal cases, using deuteron ground state and a low-energy $1^+$ bin centered at 0.37 MeV. $U_{\rm opt}^{ii'}$  has been calculated for $d+^{56}$Ni at $E_d = 32$ MeV/nucleon }
\label{local_LEP_comp_figure}
\end{figure}

Fig. \ref{local_LEP_comp_figure} shows  typical  CDCC   potentials 
  $U^{ii'}_{\rm 3N}(R)$
  for a case of $d$+$^{56}$Ni. The $U^{(00)}_{\rm 3N}$, corresponding to the deuteron ground state,   is repulsive with the strength of  5.1 MeV,  determined mostly by the UIX contact part  (Fig. \ref{local_LEP_comp_figure}a). The $2\pi$-exchange    gives a small attractive contribution with the  depth of $~1.26$ MeV coming mainly from the deuteron $d$-state. 
  Fig. \ref{local_LEP_comp_figure}b,c show  CDCC diagonal and non-diagonal 
  optical coupling potentials $U^{(ii')}_{\rm opt}(R)$  and their sums with
  $U^{(ii')}_{3N} $   for deuteron ground state and low-energy $1^+$ bin centered at 0.37 MeV.

To carry out   finite-range CDCC calculations we modified computer code FRESCO    by including the  NN potential AV18 (taken from  \cite{Wiringa}) to generate the $\phi_i$ bins. We made sure that the $n$-$p$ scattering functions published in \cite{Wiringa} were reproduced. Both $s$- and $d$-wave continuum bins    were used in $(d,p)$ and $(d,n)$ calculations. The number of bins was the same  for each component.  The bins were evenly distributed between 0 and some maximum $n$-$p$ energy which was 8.7, 46, 59 and 72 MeV for $E_d = 4.6$, 25, 32 and 40 MeV/nucleon, respectively. The corresponding bin numbers were 3, 12, 20 and 18.
Convergence with bin size and numbers was checked at all energies and   was found be the same as for any CDCC calculations, not being affected by inclusion of the 3N force. 
The $n$-$A$ and $p$-$A$ optical potentials where taken  from   widely used Koening-Delaroche global systematics \cite{KD03}.  The transferred nucleon bound-state wave functions  were obtained from  a two-body potential model of the standard geometry (radius $r_0 = 1.25$ fm and diffuseness $a = 0.65$ fm), except for $^{48}$Ca where $r_0=1.33$ fm was used.
Prior to calculations with 3N force, we  made a few standard CDCC runs with a reduced real part of the  optical potentials to simulate 3N effects. We found that even small changes, about   5$\%$, can   noticeably affect $(d,p)$ cross sections. We calculated two types of observables: angular distributions, commonly used in spectroscopic factor studies, and angle-integrated cross sections. 
The latter have been measured recently for excited final states populated in $(d,n)$ reactions via their $\gamma$-de-excitation \cite{Kah19}.
The motivation for such experiments comes from  the need in nuclear data for
nuclear astrophysics \cite{Wal81}.

\begin{figure}[t]
\includegraphics[scale=0.358]{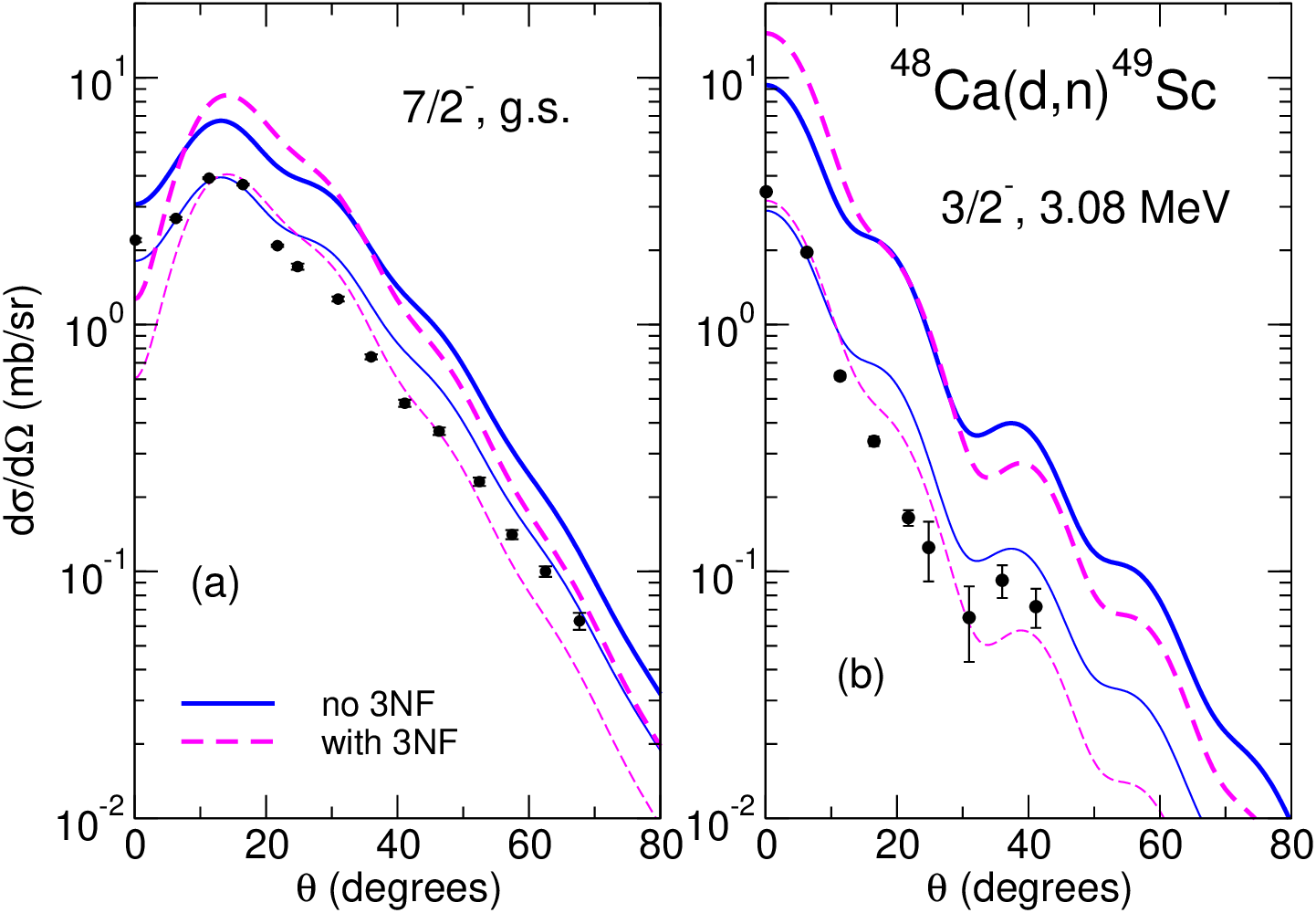}
\caption{
The   $^{48}$Ca$(d,n)^{49}$Sc    angular distributions    at $E_d=$ 40 MeV/nucleon   for ($a$) the ground   and  ($b$) first excited $^{49}$Sc states,    calculated  in CDCC with (dashed) and without (solid) $U^{(ii')}_{3N}$ assuming $S=1$ (thick lines). Thin lines show   CDCC  cross sections renormalized to  experimental data  from    \cite{Ca48}.
}
\label{Ca48_figure}
\end{figure}

We start with  $^{48}$Ca$(d,n)^{49}$Sc reaction at $E_d = 40$ MeV/nucleon,  measured in \cite{Ca48}, populating the lowest single-particle proton states   7/2$^-$ and 3/2$^-$ above the double-closed-shell stable nucleus $^{48}$Ca.
According to Independent Particle Model (IPM), their spectroscopic factors $S$ should be  one. However, accounting for NN correlations in Self-Consistent Green's Functions (SCGF) method \cite{Bar09a} and Source-Term Approach (STA) \cite{Tim11} predicts significantly  smaller values, 
$S \sim 0.5-0.7$.
The proton and neutron separation energies in $^{49}$Sc are similar and in this case the spectroscopic factors from knockout experiments are typically 45-75\% of the IPM and/or shell model predictions.
The  distorted-wave Born approximation (DWBA) and adiabatic distorted wave approximation (ADWA) analysis of    $^{48}$Ca$(d,n)^{49}$Sc, conducted in \cite{Ca48}, have also indicated  a similar quenching for $7/2^-$ state, with $S \sim 0.6$ and 0.7, respectively, and a more significant quenching for $3/2^-$, with $S \sim 0.25$ and 0.31. However, treating deuteron breakup beyond the adiabatic approximation in CDCC with the same  optical potentials (taken from Becchetti-Greenless systematics \cite{BG}) resulted in a poor match between calculated and experimental angular distributions, making spectroscopic factor extraction meaningless. We found that
CDCC with  Koening-Delaroche potentials gives a better match between predictions and the  experimental data of \cite{Ca48},  further improved by adding the 3N contribution.
 Fig. \ref{Ca48_figure}
 shows CDCC calculations with $S=1$ by thick lines while thin lines are the same calculations  multiplied by $S$ values from Table I.
 The  
 3N contribution reduces spectroscopic factors   for $^{49}$Sc(g.s.) and $^{49}$Sc$^*(3/2^-)$ by 23\% and 47\%, respectively, mainly due to a noticeable change of the  angular distribution's shape
  in the area most important for spectroscopic factor determination. 
 These spectroscopic factors  show a more significant quenching than that predicted by SCGF and STA, especially for the excited $^{49}$Sc$^*(3/2^-)$ state.

\begin{table}[t]
\begin{center}
\caption{Spectroscopic factors from CDCC analysis of angle-integrated  (for $^{56}$Ni) and   differential (for $^{48}$Ca) cross sections with and without 3N force,   in comparison with other values  from literature.}
\begin{tabular}{p{3.2cm}p{1.3cm}p{1.3 cm}p{2.3 cm}}
\hline
Reaction &   no 3NF & with 3NF & Literature  \\
\hline
$^{48}$Ca($d,n)^{49}$Sc$(g.s.)$  & 0.59 & 0.48 & 0.61$^e$; 0.72$^f$  \\
 & & (0.90)$^g$ & 0.73$^c$; 0.61$^d$ \\
$^{48}$Ca($d,n)^{49}$Sc$^*(3.08)$   & 0.31 & 0.21 & 0.25$^e$; 0.31$^f$  \\
 & & (0.39)$^g$ & 0.48$^c$; 0.55$^d$ \\
 $^{56}$Ni($d,p)^{57}$Ni$^*(0.768)$    & 0.78(22) & 0.72(21) & 0.77(31)$^a$; 0.74$^b$; 0.55$^c$; 0.68$^d$  \\
 $^{26m}$Al($d,p)^{27}$Al$^*$(0.84) & 0.10(2) &  0.09(2) & 0.08(3)$^h$; 0.11-0.13$^i$\\
 $^{26m}$Al($d,p)^{27}$Al$^*$(6.8) & 0.13(2)& 0.13(2) & 0.11(3)$^h$; 0.14$^i$ \\
\hline

\multicolumn{4}{l}{$^a$ ADWA analysis in \cite{Kah19}}\\
\multicolumn{4}{l}{$^b$ Shell model calculations  \cite{Kah19}}\\
\multicolumn{4}{l}{$^c$ Theoretical values from SCGF \cite{Bar09a,Bar09b}} \\
\multicolumn{4}{l}{$^d$ STA values recalculated here using updated harmonic }\\
\multicolumn{4}{l}{\,\,\,\,\,\,oscillator  radii as suggested in \cite{Tim21}} \\
\multicolumn{4}{l}
{$^e$ DWBA analysis in \cite{Ca48}}\\
\multicolumn{4}{l}{$^f$ ADWA analysis in \cite{Ca48}}\\
\multicolumn{4}{l}{$^g$ Becchetti-Greenless optical potential systematics} \\
\multicolumn{4}{l}{$^h$ DWBA and ADWA analysis in \cite{Alm17}} \\
\multicolumn{4}{l}{$^h$ Leading order nonlocal CDCC  analysis in \cite{Gom18}} \\
\end{tabular}

\end{center}{}
\end{table}

\begin{figure}[t]
\includegraphics[scale=0.35]{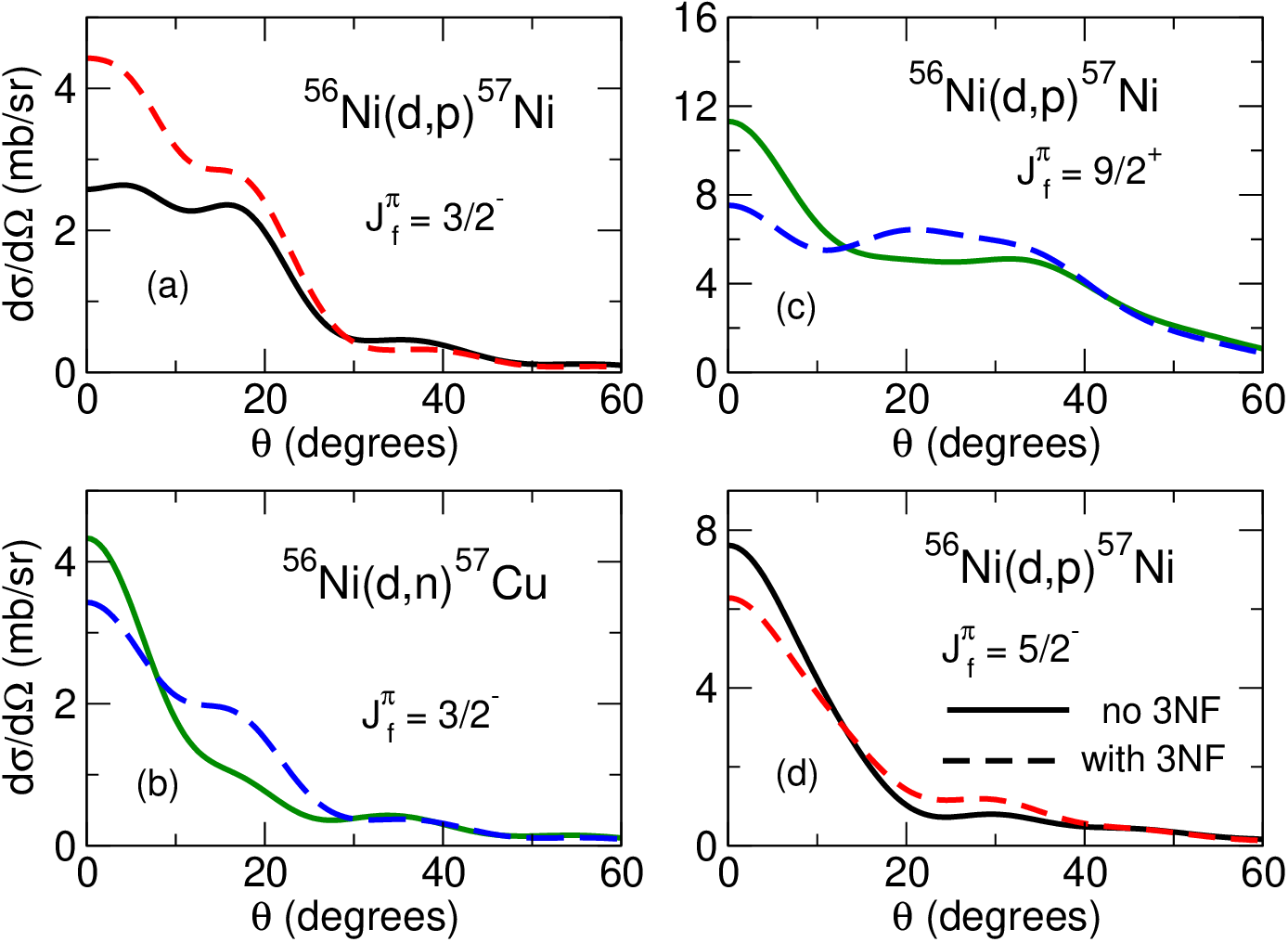}
\caption{
The CDCC  angular distributions for  $(a)$ $^{56}$Ni$(d,p)^{57}$Ni(g.s.), ($b$)  $^{56}$Ni$(d,n)^{57}$Cu(g.s.)  as well as ($c,d$)  $^{56}$Ni$(d,p)^{57}$Ni($ \tfrac{5}{2}^-,\tfrac{9}{2}^+$) reactions at  $E_d=$ 32 MeV/nucleon  obtained  with  (dashed lines) and without (solid lines)  $U^{(ii')}_{3N}$.
}
\label{Ni56_figure}
\end{figure}

We proceed with another double-closed-shell target $^{56}$Ni, close to stability but radioactive, adding   neutron or proton to it in $(d,p)$ and $(d,n)$ reactions.
The final mirror $^{57}$Ni($3/2^-$)  and $^{57}$Cu($3/2^-$) ground states  have very different valence neutron and proton separation energies, 10.248 and 0.690 MeV, respectively, and a very large difference in neutron and proton binding, +3 and -16.5 MeV. The systematic from \cite{Tos21} suggests $S \sim 1$ could be expected for $^{57}$Cu, while the $S_{\exp} = 0.58(11)$ value was derived from   neutron knockout in  \cite{Yur06}.
 The $pf$ shell model   gives $S = 0.91$ for $^{57}$Ni 
\cite{Reh98}, close to the IPM value, 
but SCGF and STA  predict significantly lower values of 0.65 \cite{Bar09b} and 0.59  \cite{Tim11}, respectively, in agreement with   knockout experiment. 
Similar values, $S = 0.67$ and 0.70,  are predicted for $^{57}$Cu both by  SCGF \cite{Bar09b} and STA \cite{Tim11}, suggesting   no strong asymmetry-dependent quenching.
When choosing these two mirror reactions
we expected  the  3N contribution would be more important in $^{57}$Ni than in $^{57}$Cu  because  rapidly decreasing  neutron  wave function outside $^{56}$Ni could amplify    contributions from  small $d$-$^{56}$Ni  separations in the entrance channel   wave function,  facilitating  the 3N interaction to occur.
However, our calculations at the energy    $E_d = 32$ MeV/nucleon, used in
  experiments with $^{56}$Ni beam  
  \cite{Kah19}, show comparable 3N effects in  mirror $^{56}$Ni$(d,p)^{57}$Ni and $^{56}$Ni$(d,n)^{57}$Cu reactions  (see Fig. \ref{Ni56_figure}a), suggesting that they will not affect the asymmetry-quenching problem. 
Depending on angles, they either increase or decrease the differential cross section so that interpretation of experimental data will depend on the angular range accessible to measurements. A similar picture occurs for $(d,p)$ reaction populating  $^{57}$Ni  excited states   $\tfrac{5}{2}^-$ at 0.768 MeV and $\tfrac{9}{2}^+$   at 3 MeV (see Fig. \ref{Ni56_figure}b). The angle-integrated cross section  $\sigma_{\rm AI} = 2.10(60)$ mb  for the first  $\tfrac{5}{2}^-$ state has been measured in \cite{Kah19}, motivated by its important role in understanding the $pr$ process in stars, with the ADWA analysis giving $S = 0.77(31)$. Based on our CDCC  results for $\sigma_{\rm AI}$, collected in Table II,  we obtained   $S = 0.78(22)$ and 0.72(21) without and with 3N force, respectively.




\begin{figure}[t]
\includegraphics[scale=0.34]{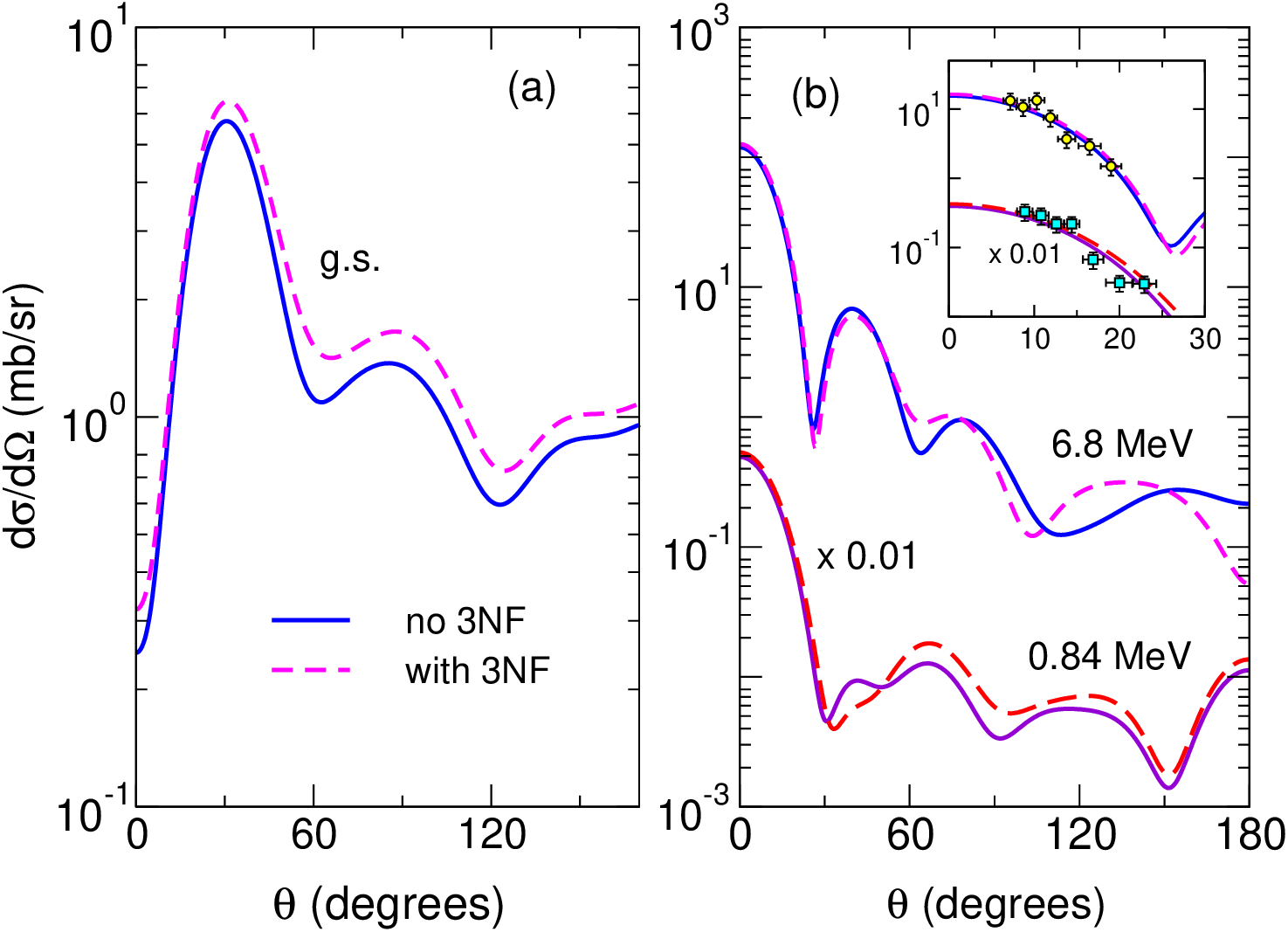}
\caption{
The CDCC  angular distributions for  $^{26m}$Al$(d,p)^{27}$Al   at $E_d= $ 4.6 MeV/nucleon   populating   ($a$) the ground  $J^{\pi}_f = 5/2^+$ and  ($b$)  excited  $J^{\pi}_f = 1/2^+$ states at $E_x = 0.84$ and 6.8 MeV,    calculated  with (dashed) and without (solid) $U^{(ii')}_{3N}$ for $S=1$. The inset shows the same calculations for $1/2^+$ state renormalised by 0.07 and 0.13 for  $E_x = 0.84$ and 6.8 MeV, respectively, in comparison with experimental data from \cite{Alm17}}
\label{Al26_figure}
\end{figure}

The two above examples were shown for deuteron incident energies in the 30-40 MeV/nucleon region assuming that 3N  effects could be especially noticeable at high deuteron energies. Many radioactive beam facilities operate at lower energies, 5-10 MeV/nucleon, where deuteron breakup can be important too \cite{Pan14}. We have assessed the 3N role  at 4.6 MeV/nucleon  for   $^{26m}$Al($d,p)^{27}$Al reaction studied in \cite{Alm17} as a surrogate for the proton capture reaction $^{26m}$Al($p,\gamma)^{27}$Si - a major destruction pathway of $^{26}$Al, in Wolf-Rayet and AGB  stars \cite{Lot09}.
Fig. \ref{Al26_figure} shows angular distributions for populating the ground $5/2^+$  state and two excited $1/2^+$ states, at $E_x = 0.84$ and 6.8 MeV. For  $l=2$ transfer to the ground state the 3N effects increase the main peak cross section by 12\%  but in the $l=0$ case this increase is 8\% and 5\% for $E_x = 0.84$ and 6.8 MeV, respectively, affecting the lowest state with a stronger bound energy more, as expected. The spectroscopic factors, collected in Table I, are slightly reduced in the 0.84 MeV case but, in general, are within limits of all previous analyses. The angle-integrated cross sections are more affected by 3N contribution, about 20\% for the lowest 1/2$^+$ state with larger separation energy while the state at 6.8 MeV is almost unaffected (see Table II).

\begin{table}[t]
\begin{center}
\caption{Angle-integrated cross sections  $\sigma_{\rm AI}$ (in mb) for reactions from first column calculated without and with $U^{(ii')}_{3N}$ assuming  $S=1$. The valence nucleon quantum numbers $lj$ and the incident  laboratory energy (in MeV/nucleon) are  in second and third columns, respectively, and the change with adding $U^{(ii')}_{3N}$ is indicated in the last column. For $(p,2p)$ reaction $E$ refers to three-body $p$+$p$+$^{24}$O final state energy. } 
\begin{tabular}{p{3.3cm}p{0.9cm}p{0.6cm}p{1.0cm}p{1.0cm}p{0.9cm}}
\hline
Reaction & $lj$ & $E_d$ & no 3NF & with 3NF & change  \\
\hline
$^{56}$Ni($d,p)^{57}$Ni$(g.s.)$    & $p_{3/2}$ & 32   &    2.002     &       2.211 & +11\% \\
$^{56}$Ni($d,n)^{57}$Cu$(g.s.)$  & $p_{3/2}$  & 32 &   1.432    &   1.699 & +19\% \\
$^{56}$Ni($d,p)^{57}$Ni$^*(0.768)$    & $f_{5/2}$ & 32    &  2.698   &      2.923 & +8\%\\
$^{56}$Ni($d,p)^{57}$Ni$^*(3.09)$    & $g_{9/2}$  & 32  & 12.70    &   12.80 & +1\% \\
$^{48}$Ca($d,n)^{49}$Sc$(g.s.)$  & $f_{7/2}$ & 40  &  6.799      &      7.376& +8\% \\
$^{48}$Ca($d,n)^{49}$Sc$^*(3.08)$   & $p_{3/2}$ & 40   & 2.211 & 2.490 & +13\% \\
$^{26m}$Al($d,p)^{27}$Al$(g.s.)$    & $d_{5/2}$ & 4.6   &   19.27    &    23.25 & +21\%  \\
$^{26m}$Al($d,p)^{27}$Al$^*(0.84 )$    & $s_{1/2}$ & 4.6   &   16.91 & 20.35 & +20\%  \\
$^{26m}$Al($d,p)^{27}$Al$^*(6.8 )$    & $s_{1/2}$ & 4.6   &   31.48   &     32.70 &  +4\%   \\
$^{24}$O($d,n)^{25}$F$(g.s.)$    & $d_{5/2}$ & 25   &          13.49     &     14.44   & +7\%                \\
$^{24}$O($d,n)^{25}$F$^*(0.495)$    & $s_{1/2}$ & 25   &         5.043        &  5.587  & +11\%                \\
$^{25}$F($p,2p)^{24}$O$(g.s.)$   & $d_{5/2}$ & 25   &     1.287 	 & 		 1.382  & +7\%      \\
$^{25}$F($p,2p)^{24}$O$(g.s.)$   & $s_{1/2}$ & 25   &     1.989 	 & 	 2.138 & +7\%        \\
\hline
\end{tabular}   
\end{center}{}
\end{table}

Finally, we consider two complementary reactions, $^{24}$O($d,n)^{25}$F and  $^{25}$F$(p,2p)^{24}$O, involving the same strongly-bound (by 14 MeV)  single-proton state above a double-closed shell  nucleus $^{24}$O with large neutron-proton asymmetry. The spectroscopic factor quenching in $^{25}$F has been reported in \cite{Tan20} where it was measured from a quasi-free $(p,2p)$ reaction at 270 MeV/nucleon  providing a value of 0.36(13), in agreement  with the STA prediction of 0.46 \cite{Tim11} within error bars. Based on many spectroscopic factor compilations from transfer reactions \cite{Kra01, Tsa05,Kay13,Man21},  one can anticipate a larger value from future $^{24}$O($d,n)^{25}$F experiments.
Here we want to check if 3N force manifestation in $(d,n)$ and $(p,2p)$ is the same or not.
To make sure that any possible differences do not come from different kinematic conditions, we carry out calculations at the same energy in the $d$+$^{24}$O and $^{24}$O+$p$+$p$ channels,  25 MeV/nucleon, available for $(d,n)$ experiments in GANIL, NSCL, JINR,  and plot the same observables - the angular distributions. The corresponding $p$+$^{25}$F energy for $(p,2p)$ experiment is 32 MeV/nucleon, available at NSCL. 
At the  energy chosen,  the $(p,2p)$ mechanism can be well described by  transfer to the $2p$ continuum  \cite{Mor15} and treated within the CDCC. 

The $U^{(ii')}_{3N}$ potentials for the $2p$-$A$ and $d$-$A$ systems are found to be similar, being repulsive and mainly determined  by the contact interaction. Unlike in the $d$-$A$ case, the $2\pi$-exchange contribution in $(p,2p)$ is   repulsive, being   negligible in the nuclear interior but  noticeable at the surface. In CDCC calculations of $(p,2p)$ reactions, we included contributions from $s$- and $p$-wave $2p$ continuum, which dominates in $(p,2p)$ reactions. 
Convergence has achieved with 15 bins for each component, evenly distributed between 0 and 41 MeV.
To check the 3N effect dependence on valence proton quantum numbers we  populate  the ground $^{25}$F$(\tfrac{5}{2}^+)$ state and not yet observed and experimentally unachievable  for $(p,2p)$ reactions excited   $^{25}$F($\tfrac{1}{2}^+$) state assuming that  $s$-wave proton energy  is the same as in $^{17}$F($\tfrac{1}{2}^+$).


\begin{figure}[b]
\includegraphics[scale=0.35]{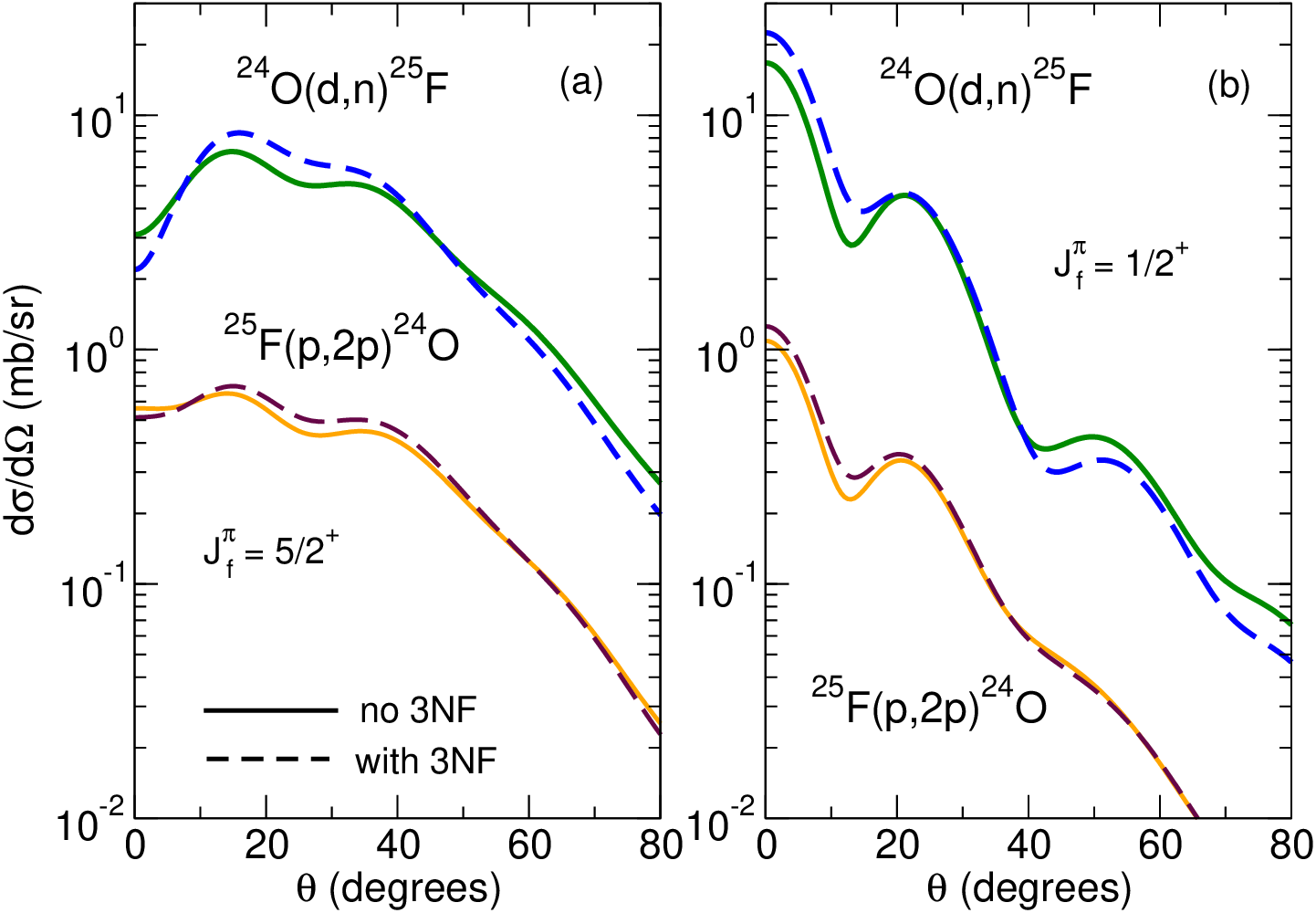}
\par
\vspace{0.3 cm}
\includegraphics[scale=0.35]{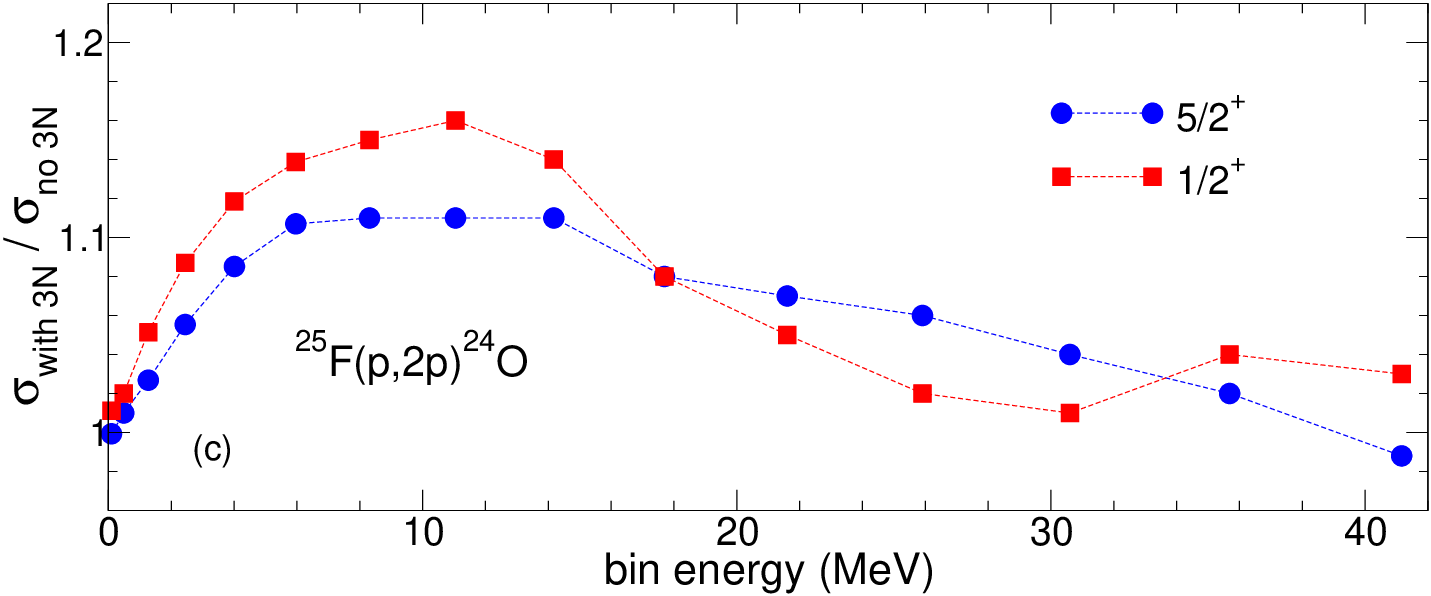}
\caption{
The CDCC  angular distributions for    $^{24}$O$(d,n)^{25}$F and $^{25}$F($p,2p)^{24}$O reactions at $E_d=E_{2p}= $ 50 MeV   populating (or knocking a nucleon from) ($a$) $^{25}$F(g.s.)  and  ($b$)  $^{25}$F$^*(1/2^+)$ states,    calculated  with (dashed) and without (solid) $U^{(ii')}_{3N}$.
Plot $(c)$ shows ratio of $\sigma_{\rm a.i.}$ for $^{25}$F($p,2p)^{24}$O reaction for each individual $2p$ bin obtained with and without $U^{(ii')}_{3N}$.}
\label{O24_figure}
\end{figure}

The  $^{24}$O$(d,n)^{25}$F and 
$^{25}$F$(p,2p)^{24}$O differential cross sections,  obtained in CDCC with and without 3N force,  are shown in Fig. \ref{O24_figure}. The 3N effects 
increase the  $(d,n)$ main peak cross sections  by 23\% and 35\% for 5/2$^+$ and 1/2$^+$, respectively, but 
in $(p,2p)$ they are less pronounced, being 8\% and 18\%. This means that the 3N-induced   difference in spectroscopic factors, obtained in complementary $(d,n)$ and $(p,2p)$ reactions from forward angles cross sections, could reach 14\%. The angle-integrated cross sections are in general less affected by the 3N force (see Table II) but this influence depends
on the $2p$ bin energy, which is illustrated in Fig. \ref{O24_figure}c for $\sigma_{\rm AI}$  calculated for transitions to individual continuum bins. 
 This means that exclusive cross section measurements corresponding to kinematically  different observations may show different 3N effects thus potentially becoming a new tool for their study. Also, in all cases 3N effects were very well simulated by a simple 5\% reduction of the real parts of proton-target optical potentials in the $2p$ channel.

Summarizing, we investigated if an additional  force  arising due to 3N interactions between neutron and proton in incoming deuteron with a target nucleon can give any effect on $(d,p)$ and $(d,n)$ reaction observables measured at 4-40 MeV/nucleon. We found that this effect is   noticeable, especially at $E > 20$ MeV/nucleon, where it can significantly alter the shape of angular distributions for $l \neq 0$ transfers. For $l=0$ transfers the forward-angle angular distributions are not much affected but their absolute values can increase up to 35 \%, with consequences for spectroscopic factors' study.
The angle-integrated cross sections, which are more sensitive to different reaction mechanisms,  are less affected by 3N force, increasing by  no more than by  20\%.  The  spectroscopic factors extracted from such observables can decrease in the same proportions. Simultaneous consideration of $^{25}$F$(p,2p)^{24}$O  and $^{24}$O$(d,n)^{25}$F reactions at the same energy in the three-body channel showed that,  without 3N contributions included in the analysis, their  spectroscopic factors  could differ by 14\%, which does contribute towards understanding the quenching-asymmetry problem. 
We must note that these conclusions have been made ignoring other uncertainties associated with   $(d,p)$, $(d,n)$ and $(p,2p)$ reactions, described for example in \cite{Aum21,Tim20b}, and using a fixed set of all reaction parameters, such as optical potentials. However, a different optical potential choice, Becchetti-Greenless, gave a similar picture here and, as shown before in \cite{Fla18,Man21} does not affect asymmetry-dependence of quenching. 
We must also note that choosing AV18+UIX interaction model   and   zero-spin targets  avoided complications due to the need for  unknown density components. The  additional $n$-$p$-$A$ force due to 3N contribution is very sensitive to the short-range parts of the NN scattering wave functions and 3N force so that further investigation involving other interaction models is needed. 
 Future work will benefit from a close collaboration with structure theorists who can  deliver missing nuclear density distributions consistent with a chosen NN+3N model. This will enable advanced single-particle spectroscopic strength studies   for any nucleus thus helping to bring nuclear structure and nuclear reactions studies together.

{\it Acknowledgments.} We are grateful Jianguo Li for providing Gamow Shell model valence nucleon densities of $^{24}$O. This work was supported by the United Kingdom Science and Technology Facilities Council (STFC) under Grants No.\ ST/L005743/1 and ST/V001108/1 and by Leverhulme Trust  Grant No.\ RPG-2019-325. M.G.R.\ acknowledges financial support by MCIN/AEI/10.13039/501100011033 under I+D+i project No.\ PID2020-114687GB-I00, by the Consejer\'{\i}a de Econom\'{\i}a, Conocimiento, Empresas y Universidad, Junta de Andaluc\'{\i}a (Spain) and ``ERDF-A Way of Making Europe'' under PAIDI 2020 project No.\ P20\_01247, and by the European Social Fund and Junta de Andalucía (PAIDI 2020) under grant number DOC-01006. 
The data underlying this article are available in the article.

\bibliographystyle{apsrev4-1}

\end{document}


\title{Single-particle spectroscopic strength  
from direct reactions with a three-nucleon force contribution. Supplement}

\author{N.K. Timofeyuk$^1$, L. Moschini$^1$ and M. G\'omez-Ramos$^2$ }
\affiliation{
$^1$Department of Physics, 
University of Surrey,
Guildford, Surrey GU2 7XH, United Kingdom \\
$^2$Departamento de FAMN, Universidad de Sevilla, Apartado 1065, 41080 Sevilla, Spain
}

\date{\today}
\maketitle

We derive formal expressions for continuum-discretized coupled-channel (CDCC) coupling potentials $U^{ii'}_{\rm 3N} (R) = \la\phi_A \phi_i| \sum_i W_{inp} | \phi_A \phi_{i'}\ra$  originating due to a 3N force taken from the Urbana-IX model \cite{Car83}. To be consistent with this model, the $n$-$p$ continuum bins $\phi_i$ must be calculated with the NN potential AV18 \cite{AV18} since  UIX   was fitted in combination with AV18 to reproduce well spectra of light nuclei \cite{Pud95}. The CDCC couplings are then used as an input to $(d,p)$, $(d,n)$ and $(p,2p)$ calculations with computer code FRESCO \cite{FRESCO}. FRESCO uses a spin-orbit angular momentum coupling scheme, which is different to the one widely adopted in few-body physics in general and in constructing AV18 in particular. The latter couples the orbital momentum $L$ of a NN pair   with the $n$-$p$ spin $S$ into a total angular momentum $J$ while FRESCO couples the $L$ with the neutron spin  first and only then adds the proton spin to get $J$. We make sure that our formalism represents correctly the $L-S$ coupling scheme and takes care of $N-N$ antisymmetrization.

\subsection{CDCC Continuum bins}

Continuum bin $\phi^i_{I m}$ with the total spin $I$ and its projection $m$ as constructed in FRESCO is
\beq
& &\phi^i_{I_d m_d}(\ve{\rm r}) = \sum_{lj_n} A^{I_d i}_{lj_n} \frac{u^i_{lj_n}(r)}{r}[[l,s_n]j_n,s_p]_{I_dm_d},
\eeqn{Fbin}
where $s_n(S_p)$ is the neutron (proton) intrinsic spin, $l$ is the orbital momentum, $\vec{j}_n = \vec{l}+\vec{s}_n$, $A$ is a normalization factor (equal to one in our case) and $u$ is the radial part which is an integral over the bin energy. The same bin can be written in the $L-S$ coupling scheme as 
\beq
\phi^i_{I_d m_d}(\ve{\rm r}) = \sum_{lS}  \frac{b^i_{lS}(r)}{r}[l,[s_n,s_p]S]_{I_dm_d} ,
\eeqn{frescobin}
where
\beq
b^i_{lS}(r) = 
\sum_{j_n} A^{I_d i}_{lj_n} u^i_{lj_n}(r) \hat{j}_n\hat{S} W(ls_nI_ds_p;j_nS),
\eeqn{bi}
$W$ is the Racah coefficient and $\hat{a} = \sqrt{2a+1}$. These bins will be used to construct CDCC coupling potentials with 3N interactions. We note that $i=0$ case correspond to deuteron bound state.

\subsection{Effective $d$-$i$ force  with 3N contact interaction}

The contact part of the Urbana-IX 3N potential is \cite{Car83}
\beq
W^{(c)}_{inp} &=& U_0 \sum _{cyc} f(r_{ni}) f(r_{ip})
\eol 
f(x) &=& t^2(x) \eol
t(x)&=&\left(1+\frac{3}{\mu x} + \frac{3}{\mu^2 x^2}\right) \frac{e^{-\mu x}}{\mu x}\left(1-e^{-b x^2}\right)^2.
\eeqn{contact}
 Here $r_{ij}$ is the distance between particles $i$ and $j$ and $cyc$ denotes cyclic permutation of the particles. We use the parameters  $U_0 = 0.0048$  MeV and $b = 2$ fm$^{-2}$ from \cite{Pud95} and $\mu$ is the pion mass.
 
 The interaction $W^{(c)}_{inp}$ can be rewritten as
\beq
W^{(c)}_{inp} 
= 4\pi \sum_{\lambda \mu}  
\,{\cal F}_{\lambda}^{(c)}(r_{di},r)\, Y^*_{\lambda \mu}(\hat{\ve{\rm r}}_{di}) Y_{\lambda \mu}(\hat{\ve{\rm r}}),
\eeqn{contact1}
where
\beq
{\cal F}_{\lambda}^{(c)}(r_{di},r) = 
\frac{1}{2} U_0 \int_{-1}^1 d\mu P_{\lambda}(\mu)
[f(x_-)f(x_+) 
\eol
+f(r)[f(x_-) +(-)^{\lambda} f(x_+)]]
\eeqn{calF}
with $x_{\pm} = \sqrt{r_{di}^2 \pm r_{di} r + \frac{1}{4}r^2}$. 
We first calculate the effective force between the $n$-$p$ pair and nucleon $i$ belonging to the target $A$ and we express it as a standard partial wave decomposition:
\beq
\la\phi^i_{I_d m_d}| W_{inp} |\phi^{i'}_{I'_d m'_d}\ra &=& \sqrt{4\pi}
\sum_{\lambda I_d I'_d} {\tilde W}_{\lambda I_d I'_d}(r_{di})
\eol
&\times &
(\lambda \mu I'_d m'_d | I_d m_d) Y^*_{\lambda \mu}(\hat{\ve{\rm r}}_{di}),
\eeqn{partial_decomposition}
where
\beq
 {\tilde W}_{\lambda I_d I'_d}(r_{di}) &=&
\sum_{ ll'j_nj'_n} B^{(\lambda)}_{lj_nI_d,l'j'_nI'_d} 
\eol
&\times&
\int_0^{\infty} dr\, A^{I_di}_{lj_n} u^{i}_{lj_n}(r){\cal F}_{\lambda}^{(c)}(r_{di},r)  A^{I'_di'}_{l'j_n'} u^{i'}_{l'j'_n}(r)   \eol
\eeqn{Wtidle}
with
 \beq
 B^{(\lambda)}_{lj_nI_d,l'j'_nI'_d}  &=&
\hat{j}_n \hat{j}'_n \hat{l}'\hat{\lambda} \hat{I}'_d(l' 0 \lambda 0 | l 0) 
\eol &\times& W(j'_n\lambda s_pI_d;j_nI'_d) W(j_n\lambda s_nl';j'_nl).
\eol
\eeqn{Bbig}
This interaction conserves isospin of the $n$-$p$  pair, which means that $(-)^{l+S} = (-)^{l'+S'}$ and because of $\delta_{SS'}$ we have $(-)^{l} = (-)^{l'}$. Therefore,   only even values of $\lambda$ will survive in the sum.

\subsection{Effective $d$-$i$ force with $2\pi$-exchange interaction}

The Urbana IX 2$\pi$ exchange potential from \cite{Car83} has contributions from commutators and anticommutators of the spin operators. 
Since we will consider 3N force contributions only for the case of targets with spin $0^+$ the contributions from anticommutators will be equal to zero. Therefore, we will consider only the following part of this potential that can be written as
\beq
W^{(2\pi)}_{inp} = 4A_{2\pi} \sum_{cyc} (\tau_i \cdot \tau_k) X_{ij}(\ve{r}_{ij})X_{kj}(\ve{r}_{kj}),
\eeqn{2piex}
where $A_{2\pi} = -0.0293$ MeV \cite{Pud95}  and
\beq
X_{ik} (\ve{r}) = [S_{ik} (\ve{r}) T(r) +  \ve{\sigma}_i\cdot \ve{\sigma}_k ]Y(r)
\eeqn{Xdef}
is the coordinate-space pion propagator that contains the tensor operator
\beq
S_{ik} (\ve{r}) = 3(\ve{\sigma}_i \cdot \hat{\ve{r}}) (\ve{\sigma}_k \cdot \hat{\ve{r}} ) -  \ve{\sigma}_i\cdot \ve{\sigma}_k.
\eeqn{Sdef}
The functions in Eq. (\ref{Xdef}) are
\beq
Y(r) &=& \frac{e^{-\mu r}}{\mu r}\left(1-e^{-2r^2}\right), \label{Y}\\
T(r) &=& \left(1 + \frac{3}{\mu r}+ \frac{3}{(\mu r)^2}\right)\left(1-e^{-2r^2}\right) \label{T}.
\eeqn{YT}
 The matrix element $\la\phi^i_{I_d m_d}| W_{inp}^{(2\pi)} |\phi^{i'}_{I'_d m'_d}\ra $ could be represented by a partial decomposition, similar to Eq. (\ref{partial_decomposition}), with the radial part given by
 \beq
 \tilde W^{ii'}_{\lambda I'_d I_d}(r_{dj}) =
 \sum_{j_nj'_nll'}  \int_0^{\infty} dr \tilde u^{(i)}_{\alpha}(r) {\cal F}^{(2\pi)}_{\lambda \alpha \alpha'}(r_{di},r)
\tilde u^{(i')}_{\alpha'}(r) ,
\eol
\eeqn{W2pi}
where we use shorter notations, $\alpha \equiv \{lj_nI_d\}$ and $\tilde u^{(i)}_{\alpha}(r)= A^{I_d i}_{lj_n}u^i_{lj_n}(r)$.   The $ \tilde W^{ii'}_{\lambda I'_d I_d}(r_{dj})$ contains formfactor
\beq
 {\cal F}^{(2\pi)}_{\lambda \alpha \alpha'}(r_{di},r)=
 -3 B^{(\lambda)}_{\alpha\alpha'}
{\cal H}_{\lambda}^{(0)}(r_{di},r) \,\,\,\,\,\,\,\,\,\,\,\,\,\,\,\,\,\,\,\,\,\,\,\,\,\,\,\,\,\,\,\,\,\,\,\,\,\,\,\,\,\,\,\,\,\,\,\,\,\,\,\,\,\,\,\,\,\,\,\,\,\,
\eol 
 + \sum_{\lambda'}\left[ {\cal C}^{(1)}_{\lambda \lambda' \alpha\alpha'}\left( r^2_{dj}  {\cal H}_{\lambda'}^{(1)}(r_{di},r) +r^2  {\cal H}_{\lambda}^{(3)}(r_{di},r) \right) \right.
\eol
+\left. {\cal C}^{(2,0)}_{\lambda \lambda' \alpha\alpha'} rr_{dj} {\cal H}_{\lambda'}^{(2,0)}(r_{di},r)
+{\cal C}^{(2,1)}_{\lambda \lambda' \alpha\alpha'} rr_{dj} {\cal H}_{\lambda'}^{(2,1)}(r_{di},r) \right], \eol
 \eeqn{}
with coefficients ${\cal C}$ given by the following equations:
 \beq
 {\cal C}^{(1)}_{\lambda \lambda' \alpha\alpha'} &=&
  \sum_{  \tilde \lambda }  6 \hat{\lambda'} \hat{\lambda}\hat{l}'\hat{\tilde \lambda} \hat{I}'_d  (1 0 1 0 | {\tilde \lambda} 0) (\tilde \lambda 0 \lambda 0 | \lambda' 0) 
 \eol &\times& (l' 0 \lambda' 0 | l 0)
{\cal S}_{\lambda \tilde \lambda \lambda' \alpha \alpha' },
 \eeqn{C1}
\beq
{\cal C}^{(2,0)}_{\lambda \lambda' \alpha\alpha'} =
\sum_{  l_1   \tilde \lambda } 
(-)^{\lambda'+\lambda } 6
 \hat{\lambda}'^2 \hat{\tilde \lambda}^2 \hat{l}' \hat{I}'_d  \hat{l}_1(\lambda 1 l_1 1; \lambda' \tilde \lambda)  
 \eol \times
(1 0 \lambda' 0 | l_1 0) 
(1 0 \lambda' 0 | \lambda 0) (l' 0 l_1 0 | l 0) {\cal S}_{\lambda \tilde \lambda l_1 \alpha \alpha' },
\eeqn{C20}
\beq
{\cal C}^{(2,1)}_{\lambda \lambda' \alpha\alpha'} =
\sum_{  l_1   \tilde \lambda } 
(-)^{\lambda'+\lambda +\tilde \lambda} 6
 \hat{\lambda}'^2  \hat{\tilde \lambda}^2 \hat{l}' \hat{l}_1\hat{I}'_d (\lambda 1 l_1 1; \lambda' \tilde \lambda)  
 \eol \times
(1 0 \lambda' 0 | l_1 0) 
(1 0 \lambda' 0 | \lambda 0) (l' 0 l_1 0 | l 0) {\cal S}_{\lambda \tilde \lambda l_1 \alpha \alpha' },
\eeqn{C21}
where
\beq
 {\cal S}_{\lambda \tilde \lambda l_1 \alpha \alpha' } =\sum_{SS'}  (-)^{I'_d + I_d + \lambda }t_{lSl'S'}  \hat{j}_n\hat{j}'_n\hat{S}^2\hat{S}'^2
W(ls_nI_ds_p;j_nS)
\eol \times
 W(l's_nI'_ds_p;j'_nS')
\left\{\begin{array}{lll} 
{\dem}& {\dem} & {S} \\ {\dem}& {\dem} & {S'} \\ {1} & {1} & { \tilde\lambda} \end{array}\right\}
\left\{\begin{array}{lll} 
{l'}& {S'} & {I'_d}\\ {l_1} & {\tilde \lambda} & { \lambda} \\ {l}& {S} & {I_d}  \end{array}\right\}. \eol
\eeqn{}
The antsymmetrization is cared for by coefficients $t_{lSl'S'} $ which are 
\beq
t_{lSl'S'} &=& 0, \,\,\,\,\, (-)^{l+S} \neq (-) ^{l'+S'} \eol
t_{lSl'S'} &=& 1, \,\,\,\,\, (-)^{l+S} = (-) ^{l'+S'} = 1\eol
t_{lSl'S'} &=& -3, \,\,\,\,\, (-)^{l+S} = (-) ^{l'+S'} = -1
\eeqn{}
The expressions for ${\cal H}$ are the following:
\beq
{\cal H}^{(\gamma)}_{\lambda}(r_{di},r)=  2A_{2\pi}
\int _{-1}^1 d\mu P_{\lambda}(\mu) Y_-Y_+{\cal G}^{(\gamma)}(r_{di},r,\mu), \eol
\eeqn{H}
where
\beq
{\cal G}^{(0)}(r_{di},r,\mu)& =& (1-T_-)(1-T_+), \eol
{\cal G}^{(1)}(r_{di},r,\mu) &=& {\cal T} + {\cal X}_1+ {\cal X}_2, \eol
{\cal G}^{(1)}(r_{di},r,\mu) &=& \frac{1}{2}\left(-{\cal T} - {\cal X}_1+ {\cal X}_2\right), \eol
{\cal G}^{(2,0)}(r_{di},r,\mu) &=& \frac{1}{2}\left({\cal T} - {\cal X}_1+ {\cal X}_2\right),\eol
{\cal G}^{(3)}(r_{di},r,\mu) &=& \frac{1}{4}\left({\cal T} +{\cal X}_1+ {\cal X}_2\right)
\eeqn{Gs}
with
\beq
{\cal T} &=& 9 \left(r^2_{dj} - \frac{1}{4} r^2\right)  \frac{ T_{-} T_{+} }{x_{-}^2 x_{+}^2}, \eol
{\cal X}_1 &=& \frac{3T_{-}(1 - T_{+})}{x_-^2}, \eol
{\cal X}_2 &=& \frac{3T_{+}(1 - T_{-})}{x_+^2}
\eeqn{calTX}
and $Y_- = Y(x_-)$, $Y_+= Y(x_+)$, $T_- = T(x_-)$ and $T_+ = T(x_+)$ with $Y$ and $T$ given by Eqs.    (\ref{Y}) and (\ref{T}), respectively.

\subsection{CDCC couplings for 3N force}

The CDCC as implemented in FRESCO requires matrix elements of the operators that couple states
\beq
| (L I_d)J,I_t;J_T\ra =\left[ \left[Y_L(\hat{\ve{R}}) \otimes \phi^i_{I_d}(\ve{r})\right]_J \otimes \phi_{I_t} \right]_{J_T}
\eeqn{cdcc-state}
with well-defined relative orbital momentum $L$ between the target  and the center of mass of bin $i$, its spin $ I_d$  and the target spin $I_t$ that are coupled into the total angular momentum $J_T$. In the case when $I_t = I'_t = 0$ and  potential $W_{inp}$ does not involve spin and isospin operators acting on target nucleons one can show that
\beq
\la (L I_d)J,I_t;J_T|W_{inp} | (L' I'_d)J',I'_t;J'_T\ra
=
 \delta_{JJ_T} \delta_{J'J'_T} 
 \eol \times \sum_{\lambda } 4\pi 
\hat{I}'_d  \hat{L}' \hat{\lambda}(L'0 \lambda 0 | L 0)(-)^{I'_d+ J_T +L}
\left\{ \begin{array}{lll} 
{I'_d}& {\lambda} & {I_d} \\ {L} & {J_T} & { L'} \end{array}\right\}
\eol \times \int_0^{\infty} dr_{di} r_{di}^2\rho_{\lambda}(r_{di},R){\tilde W}_{\lambda I_dI'_d}(r_{di}),\,\,\,\,\,\,\,\,\,\,\,\,\,\,\,\,
\eeqn{WinpME}
where ${\tilde W}_{\lambda I_dI'_d}(r_{di})$ is the effective transition interaction connecting initial and final bin states with a target nucleon,      discussed in the two sections above, and 
\beq
\rho_{\lambda}(r_{di},R) 
  = \frac{1}{2} \int_{-1}^1 d\mu \, P_{\lambda}(\mu) \rho_A \left( \left|\ve{r}_{di} - \ve{R} \right| \right). \,\,\,\,\,\,\,\,
\eeqn{}
Here $\rho_A$ is the matter density distribution in the target $A$.
The FRESCO convention for coupling potentials in the case of the $I_t = I_t = 0$ targets is
\beq
\la (L I_d)J,I_t;J_T|U_{\lambda} | (L' I'_d)J',I'_t;J'_T\ra
= (-)^{I_d + L + J_T} \,\,\,\,\,\,\,\,\,\,\,\,\,\,\,\,\,\,
\eol
\times
\hat{L}' (L'0 \lambda 0 | L 0) \left\{ \begin{array}{lll} 
{I'_d}& {\lambda} & {I_d} \\ {L} & {J_T} & { L'} \end{array}\right\}U^{\rm FRESCO}_{\lambda I_d I'_d}(R), \,\,\,\,\,\,\,\,\,\,\,\,\,\,\,\,\,\,
\eeqn{fresco}
Comparing it with Eq. (\ref{WinpME}) we conclude that
\beq
U^{\rm FRESCO}_{\lambda I_d I'_d}(R)
&=& (-)^{I_d-I'_d}4\pi \hat{\lambda} \hat{I}'_d 
\eol &\times& \int_0^{\infty} dr_{di} r_{di}^2\rho_{\lambda}(r_{di},R){\tilde W}_{\lambda I_dI'_d}(r_{di}).
\eol
\eeqn{Ufresco}

\bibliographystyle{apsrev4-1}